\documentclass[aps, prb, reprint, superscriptaddress, floatfix,]{revtex4-1}
\usepackage{amsmath, amssymb, graphicx}
\usepackage[hidelinks]{hyperref}

\begin{document}

\title{Topological superconductivity in Kondo-Kitaev model}

\author{Wonjune Choi}
\affiliation{Department of Physics, University of Toronto, Toronto,
  Ontario M5S 1A7, Canada}
\author{Philipp W. Klein}
\affiliation{Institut f\"{u}r Theoretische Physik, Universit\"{a}t zu K\"{o}ln, Z\"{u}lpicher Stra\ss e 77, K\"{o}ln 50937, Germany}
\author{Achim Rosch}
\affiliation{Institut f\"{u}r Theoretische Physik, Universit\"{a}t zu K\"{o}ln, Z\"{u}lpicher Stra\ss e 77, K\"{o}ln 50937, Germany}
\author{Yong Baek Kim}
\affiliation{Department of Physics, University of Toronto, Toronto,
  Ontario M5S 1A7, Canada}
\affiliation{Canadian Institute for Advanced Research, Toronto, Ontario M5G 1Z8, Canada}
\affiliation{School of Physics, Korea Institute for Advanced Study, Seoul 130-722, Korea}

\begin{abstract}
We investigate possible topological superconductivity in the Kondo-Kitaev model on the honeycomb lattice, where the Kitaev spin liquid 
is coupled to conduction electrons via the Kondo coupling. We use the self-consistent Abrikosov-fermion mean-field theory to map out
the phase diagram. Upon increasing the Kondo coupling, a first order transition occurs from the decoupled phase of spin liquid and conduction electrons 
to a ferromagnetic topological superconductor of Class D with a single chiral Majorana edge mode. This is followed by a second order transition into a paramagnetic
topological superconductor of Class DIII with a single helical Majorana edge mode. 
These findings offer a novel route to topological superconductivity in the Kondo lattice system.
We discuss the connection between topological nature of the Kitaev spin liquid and topological superconductors obtained in this model.

\end{abstract}

\maketitle

\section{Introduction} \label{sec:intro}

Quantum spin liquid with Ising topological order can be regarded as the infinite on-site repulsion limit of an underlying superconductor\cite{Senthil_Z2gauge, Balents_QSLintro}.
In other words, it is a ``projected" superconductor, where the electrons live in the constrained Hilbert space with exactly one electron per site and therefore with no charge transport.
The fractionalized charge-neutral excitations in such systems have built-in pairing correlation with emergent Ising gauge structure.
Hence, introducing charge fluctuations in such a quantum spin liquid phase is a promising route to obtain unconventional superconductivity.
This may be achieved by doping the spin liquid\cite{Anderson_RVB, Senthil_dopedU1, You_dopedJK, Schmidt_dopedJKG}, pushing the system towards the metal-insulator transition\cite{Motrunich_nearMott}, and using the Kondo-coupling to itinerant electrons\cite{Senthil_FL*, Seifert_KondoKitaev}.

In this work, we explore the emergence of topological superconductivity in the Kondo-Kitaev model, where the Kitaev quantum spin liquid is Kondo-coupled to conduction electrons on the two-dimensional honeycomb lattice, using the slave-particle self-consistent mean-field theory. 
The Kitaev spin liquid is the ground state of the exactly solvable Kitaev model of local moments with bond-dependent Ising interactions on the honeycomb lattice. It has Ising or $\mathbb{Z}_2$ topological order and supports charge-neutral Majorana fermion excitations with Dirac dispersion\cite{Kitaev_Honeycomb}.
The purpose of this study is to understand what kinds of unconventional superconductors are possible and how the topological nature of the Kitaev spin liquid may be manifested in the resulting superconducting state. 

The main finding of the current work is the identification of two kinds of emergent topological superconductors\cite{Sato_TSCreview}, when the Kondo coupling is sufficiently large. 
For this purpose, we focus on the model where the Kitaev coupling to the local moments is significantly larger than the hopping amplitude or the bandwidth of the conduction electrons.
It is clear that one obtains the decoupled Kitaev-spin-liquid and conduction-electron system when the Kondo coupling is small, as any short range interaction would be an irrelevant perturbation because of the vanishing density of states of Majorana fermions in the Kitaev spin liquid.
This state is an example of the so-called FL$^*$ or fractionalized Fermi liquid phase\cite{Senthil_FL*, Senthil_weakMagnetism}.
We find that, upon increasing the Kondo coupling, the Kondo hybridization between local moments and conduction electrons becomes finite via a first order transition and the system enters a superconducting state. This superconducting state is a ferromagnetic (time-reversal breaking) topological superconductor (FM-TSC) of Class D in the ten-fold way classification scheme\cite{Schnyder_tenfoldway}, with a single chiral Majorana edge mode. Upon increasing the Kondo interaction further, there exists a second order phase transition to a paramagnetic (time-reversal preserving) topological superconductor (PM-TSC) of Class DIII\cite{Schnyder_tenfoldway}, with a single helical Majorana edge mode. 

These results may be heuristically understood as follows.
The Kitaev spin liquid is described by the projective symmetry group of Ising variety or $\mathbb{Z}_2$ invariant gauge group, which is the origin of Ising topological order\cite{Kitaev_Honeycomb}.
The Dirac dispersion of fractionalized Majorana fermions is protected by projective time-reversal and particle-hole symmetry, and hence the Majorana representation of the Kitaev model can be cast as the Class BDI Hamiltonian of the Majorana fermions\cite{Kitaev_Honeycomb, Trebst_gaplessKitaev, Schaffer_JKmodel}.
When time-reversal symmetry is broken, for example, by an external magnetic field, the spectrum of Majorana fermions becomes gapped and the ground state is a chiral spin liquid with Chern number $\pm 1$, which can be described by the Class D representation of the Majorana fermions.

When the Kondo hybridization with conduction electrons becomes finite, the local moments are mixed with Ising-gauge-neutral conduction electrons so that the composite system is no longer invariant under the Ising gauge fluctuations of Majorana fermions.
Thus, the emergent Ising gauge structure (or the projective symmetry group structure) is lost\cite{Senthil_FL*, SCfromFL*} and the system may spontaneously break the symmetries that protect
the spin liquid in the decoupled limit\cite{Rau_2014}.
This is the way that the FM-TSC of Class D appears via time-reversal symmetry breaking with finite Kondo hybridization. 
Notice that this FM-TSC inherits the same topological property of the time-reversal symmetry breaking chiral spin liquid mentioned earlier.
Since the small Kondo coupling is irrelevant, the transition from FL$^*$ to FM-TSC occurs at a sufficiently large Kondo coupling.
When the Kondo coupling is further increased, a continuous transition to a time-reversal symmetry preserving PM-TSC occurs.
All mean-field amplitudes which break time-reversal symmetry vanish in this state.

Our results presented here are somewhat different from a previous study using a Majorana-based mean-field approach\cite{Seifert_KondoKitaev}, where a non-topological gapless superconductor was found.
The main reason for the different results is that we keep all possible mean-field channels (92 real-valued mean-field parameters) in our Abrikosov fermion mean-field theory. Choosing a subset of these parameters would correspond to the previous work.

A number of possible candidate topological superconductors such as Cu-doped Bi$_2$Se$_3$\cite{Ando_Bi2Se3} and Sr$_2$RuO$_4$\cite{Maeno_Sr2RuO4} have been extensively studied, and recently several candidate materials for the Kitaev spin liquid\cite{JKmechanism, KitaevMagnetReview, Trebst_2017, Rau_review, William_review, Takagi_H3LiIr2O6} have been identified.
It is our hope that our work would shed light on a new route to topological superconductors, possibly starting from such Kitaev-like materials.

The remainder of the paper is organized as follows. 
In Sec.~\ref{sec:model}, we introduce the Kondo-Kitaev model and briefly summarize its symmetry properties. In Sec.~\ref{sec:MFT}, we construct the most general nearest-neighbour complex fermion (Abrikosov fermion) mean-field Hamiltonian that can reproduce the exact ground state energy and excitation spectrum in the Kitaev limit.
Sec.~\ref{sec:FLstar} reviews the projective symmetries of the Kitaev model and discusses how the projective time-reversal symmetry protects the gapless Majorana cones.
In Sec.~\ref{sec:SC}, we present our main result on topological superconductors via the Kondo hybridization.
We first examine which symmetry is spontaneously broken with finite hybridization and identify the symmetry class of the superconductors using the ten-fold way classification of topological insulators and superconductors\cite{Schnyder_tenfoldway, Kitaev_tenfoldway, Chiu_tenfoldway}.
From the bulk topological invariant and the non-trivial edge mode, we analyze the topological property of the hybridized topological superconductors and discuss the connection to the topological nature of the Kitaev spin liquid.
We summarize and conclude our work in Sec.~\ref{sec:conc}.

\section{The Kondo-Kitaev Model} \label{sec:model}

\begin{figure}[t]
\centering
\includegraphics[width=0.5\textwidth]{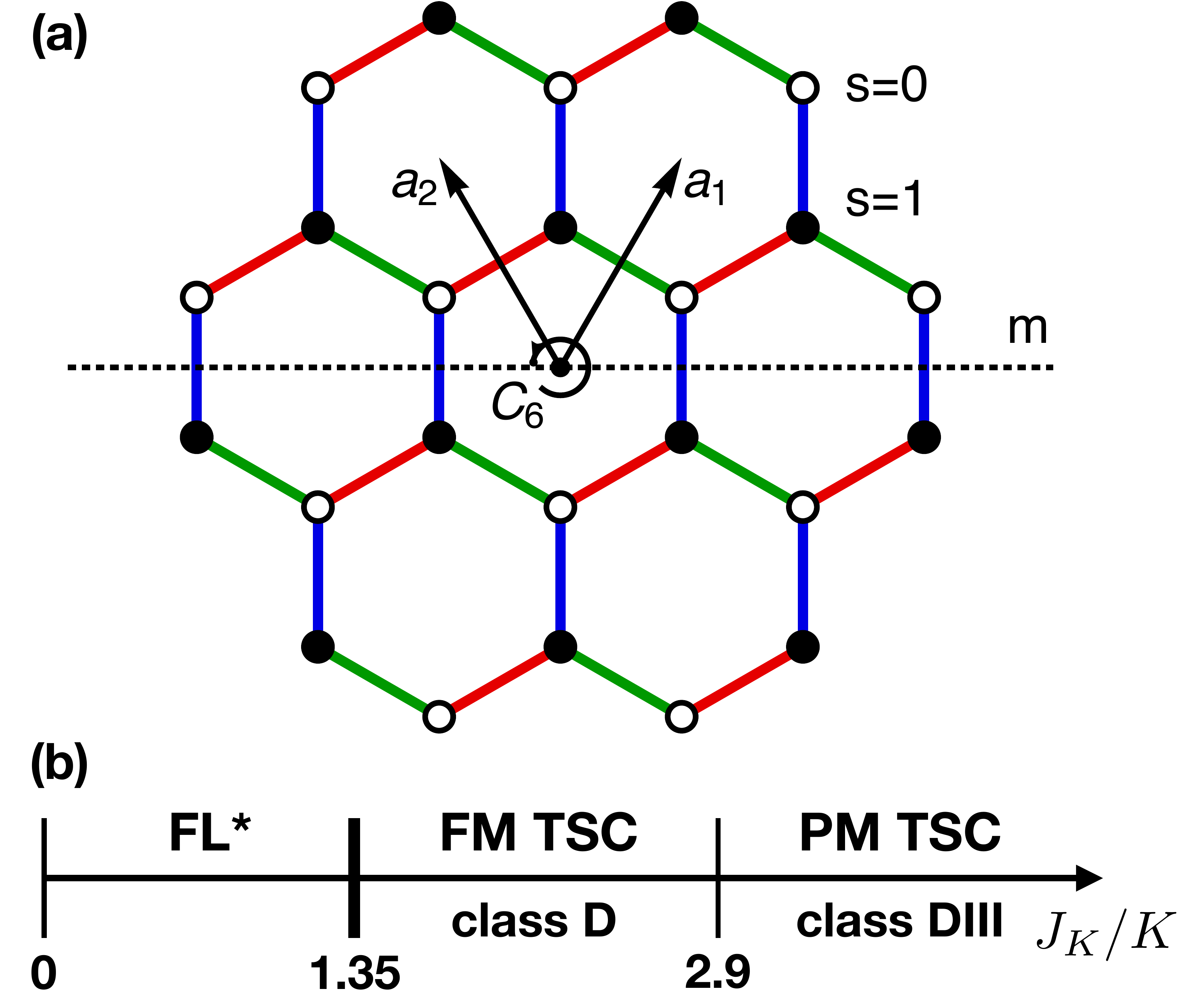}
\caption{(a) Symmetries of the Kondo-Kitaev model: time-revsersal $\mathcal{T}$, lattice translations $T_{1,2}$, mirror reflection $m$ and six-fold rotations $C_6$. Red, green, and blue links denote the $x\text{-}, y\text{-}, z\text{-}$links of the Kitaev interactions, respectively. \\
(b) Phase diagram with $t/K =0.2$ and $n_c = 0.7$.
Strong Kondo coupling can hybridize electrons and spins, and it drives discontinuous transition from a fractionalized Fermi liquid (FL*) into a ferromagnetic chiral topological superconductor (FM-TSC).
Further increase of the Kondo coupling yields continuous topological transition into a paramagnetic topological superconductor (PM-TSC) with time-reversal symmetry.}
\label{fig:lattice}
\end{figure}

\subsection{Model Hamiltonian} \label{ssec:KondoKitaev}
We start with the following Kondo lattice model on the honeycomb lattice.
The conduction electrons $c_{i\alpha}$ are described by the nearest-neighbour tight-binding model with the chemical potential $\mu$ controlling the electron filling fraction $n_c = \langle c_{i\alpha}^\dagger c_{i\alpha} \rangle $.
We consider the ferromagnetic link-dependent Kitaev interaction between local moments $\vec{S}_i$, which are coupled to conduction electrons via the on-site antiferromagnetic Heisenberg Kondo coupling. The model Hamiltonian is $H = H_c + H_K + H_{\rm Kondo}$ with
\begin{align}
&H_c = -t  \sum_{\langle ij \rangle} \left(c_{i\alpha}^\dagger c_{j\alpha} + c_{j\alpha}^\dagger c_{i\alpha} \right) -\mu \sum_i c_{i\alpha}^\dagger c_{i\alpha} \ , \label{eq:Hc} \\
&H_K = -K \sum_{a\text{-link}} S_i^a S_j^a \ ,
\label{eq:HK}
\\
&H_{\rm Kondo} =\frac{J_{K}}{2}  \sum_{i} (c_{i \alpha}^\dagger \vec{\tau}_{\alpha\beta} c_{i\beta}) \cdot \vec{S}_i \ ,
\label{eq:Hv}
\end{align}
where $\tau^a$ are the Pauli matrices.
We are assuming the Einstein summation convention for the repeated Greek indices (e.g., $\alpha = \uparrow, \downarrow$), but not for the Latin indices.

In order to map out the phase diagram of the Kondo-Kitaev model [FIG.~\ref{fig:lattice} (b)], we focus on the case of $t/K = 0.2$ and electron filling fraction $n_c = 0.7$.
We have checked that other similar choices of $t/K$ and $n_c$ only change the positions of the phase boundaries, but do not alter the nature of each of the phases.

\subsection{Symmetries} \label{ssec:symm}

The Hamiltonian $H$ is invariant under time-reversal, $\mathcal{T}$, and space group symmetries of the honeycomb lattice including lattice translations $T_{1,2}$ along $\mathbf{a}_{1,2}$, six-fold rotations $C_6$ around the centre of the hexagon, and mirror reflection $m$ [FIG.~\ref{fig:lattice} (a)].
Due to the spin-orbit coupled nature of the Kitaev interaction, $SU(2)$ spin rotational symmetry is explicitly broken. Thus, the spin should be rotated accordingly upon mirror reflection or spatial rotation\cite{You_dopedJK}, i.e., under the symmetry transformation $g$
\begin{equation}
c_{i\sigma} \rightarrow (R_g^\dagger)_{\sigma\sigma'} c_{g(i)\sigma'}, \quad S_i^a \rightarrow O^{ab} S^b_{g(i)},
\end{equation}
where
\begin{align}
&R_{\mathcal{T}} = -i\tau^y \mathcal{K}, \quad \mathcal{K} i \mathcal{K} = -i \label{eq:UT} \\
&R_{T_1} = R_{T_2} = \tau^0, \label{eq:UT12}\\
&R_m = \frac{i}{\sqrt{2}}(\tau^x + \tau^y), \label{eq:Um} \\
&R_{C_6} = \frac{1}{2}(\tau^0 + i\tau^x+i\tau^y+i\tau^z), \label{eq:UC6}
\end{align}
and
\begin{align}
&O_{\mathcal{T}} = -  I_{3\times3},\\
&O_{T_1}=O_{T_2} = I_{3\times3},\\
&O_m =
\begin{pmatrix}
0 & 1 & 0 \\
1 & 0 & 0 \\
0 & 0 & -1
\end{pmatrix}, \\
&O_{C_6} =
\begin{pmatrix}
0 & 0 & 1 \\
1 & 0 & 0 \\
0 & 1 & 0
\end{pmatrix}.
\end{align}

\section{Abrikosov fermion mean-field theory} \label{sec:MFT}

\subsection{The Kitaev model} \label{ssec:Kitaev}
\subsubsection{Spin liquid channel}

The Kitaev model $H_K$ can be solved exactly\cite{Kitaev_Honeycomb} by representing the spin operator $S_i^a =\frac{i}{2} \gamma_i^a \gamma_i^0$ with the Majorana fermions $\{ \gamma_i^\mu, \gamma_j^\nu \} = 2\delta_{ij} \delta^{\mu\nu}$ and the constraint $\gamma_i^x \gamma_i^y \gamma_i^z \gamma_i^0 = 1$ at every site $i$.
We first consider the Majorana fermion mean-field theory of $H_K$:
\begin{multline}
H_K^{MF} = \frac{K}{4} \sum_{a\text{-link}}
\langle i \gamma_i^a \gamma_j^a \rangle  i \gamma_i^0 \gamma_j^0 + \langle i \gamma_i^0 \gamma_j^0 \rangle  i \gamma_i^a \gamma_j^a \\
-\langle i \gamma_i^a \gamma_j^a \rangle \langle  i \gamma_i^0 \gamma_j^0 \rangle.
\label{eq:HKmf}
\end{multline}
The exact spectrum of the gapless Majorana fermions $\gamma^0$ and the exact ground state energy $E_0 = -0.3936 K$ (per unit cell) can be reproduced with the self-consistent mean-field parameters\cite{You_dopedJK},
\begin{equation}
\langle i \gamma_i^a \gamma_j^a \rangle = 1, \quad \langle i \gamma_i^0 \gamma_j^0 \rangle = -0.524864.
\label{eq:Majsol}
\end{equation}

The Majorana fermion mean-field theory can be transformed to the complex (Abrikosov) fermion mean-field theory with the unitary transformation $f_\uparrow = \frac{1}{2} \left(\gamma^0 - i \gamma^z \right)$ and $f_\downarrow = \frac{1}{2} \left( \gamma^y - i \gamma^x \right)$.
The change of basis reduces the constraint $\gamma_i^x \gamma_i^y \gamma_i^z \gamma_i^0 = 1$ into $f_{i\alpha}^\dagger f_{i\alpha} = 1$ and the spin operator into the familiar expression $S_i^a = \frac{1}{2} f_{i\alpha}^\dagger \tau^a_{\alpha\beta}f_{i\beta}$ with help of the one-spinon-per-site constraint\cite{Burnell_KitaevMFT}.
The mean-field Hamiltonian becomes the Bogoliubov-de Gennes (BdG) Hamiltonian with the singlet and triplet hopping (and pairing).
Without loss of generality, let us focus on the mean-field Hamiltonian for the $z$-link.
For the compact expression, we introduce the link variables
\begin{align}
&\hat{\chi}_{ij}  = f_{i\sigma}^\dagger f_{j\sigma}, &
&\hat{\eta}_{ij}  = f_{i\alpha} (-i\tau^y)_{\alpha\beta} f_{j\beta}, \\
&\hat{E}_{ij}^a =  f_{i\alpha}^\dagger \tau^a_{\alpha\beta} f_{j\beta}, &
&\hat{D}_{ij}^a =  f_{i\alpha} (i\tau^y \tau^a)_{\alpha\beta}f_{j\beta}.
\end{align}
Then $H_K^{MF}$ in terms of $f$ fermions is written as
\begin{multline}
\left(H_K^{MF}\right)_z = \frac{K}{2} \sum_{z\text{-link}} \left ( \chi_{ij}^* \hat{\chi}_{ij} + E_{ij}^{z*}\hat{E}_{ij}^z \right. \\
\left. \qquad \qquad \qquad -D_{ij}^{x*} \hat{D}_{ij}^x  - D_{ij}^{y*} \hat{D}_{ij}^y + \mathrm{h.c.} \right)\\
-\left( |\chi_{ij}|^2 + |E_{ij}^z|^2 - |D_{ij}^x|^2 - |D_{ij}^{y} |^2 \right),
\label{eq:HKzmf}
\end{multline}
where
\begin{align}
\chi_{ij} &= \langle \hat{\chi}_{ij} \rangle = E^z_{ij} = \langle \hat{E}^z_{ij} \rangle
= -\frac{i}{4}  \left( \langle i \gamma_i^0 \gamma_j^0 \rangle + \langle i\gamma_i^z \gamma_j^z \rangle \right) \nonumber \\
&= -0.118784 i, \label{eq:chiz} \\
D^x_{ij} &= \langle \hat{D}^x_{ij} \rangle = -iD^y_{ij} =-i \langle \hat{D}^y_{ij} \rangle = -\frac{i}{4}\left( \langle i \gamma_i^0 \gamma_j^0 \rangle - \langle i\gamma_i^z \gamma_j^z \rangle\right) \nonumber \\
&= 0.381216i, \label{eq:Dxz}
\end{align}
and the other link variables have vanishing expectation values.
Note that the mean-field parameters (the expectation values of the link variables) can be deduced not only from Eq.~(\ref{eq:Majsol}) with the change of basis, but also from the self-consistency conditions in Eqs.~(\ref{eq:chiz}) and (\ref{eq:Dxz}).
Because Eq.~(\ref{eq:HKzmf}) is unitary equivalent to Eq.~(\ref{eq:HKmf}), it reproduces the exact excitation spectrum and the ground state energy by construction.

In an alternative route, since the ground state wavefunction of any mean-field Hamiltonian is a single Slater determinant state, we can compute $\langle H_K \rangle$ directly for
the ground state of a complex fermion mean-field theory after we represent the spin as $S_i^a = \frac{1}{2} f_{i\alpha}^\dagger \tau^a_{\alpha\beta}f_{i\beta}$.
Using Wick's theorem, we obtain
\begin{align}
&\langle H_K\rangle  = -\frac{K}{4}\sum_{a\text{-link}} \langle (f_{i\alpha}^\dagger \tau^a_{\alpha\beta}f_{i\beta})(f_{j\alpha}^\dagger \tau^a_{\alpha\beta}f_{j\beta}) \rangle \nonumber \\
&= \sum_{a\text{-link}} \frac{K}{8} \left( |\chi_{ij}|^2 + |E_{ij}^a|^2 - |E_{ij}^b|^2 - |E_{ij}^{c} |^2 \right) \nonumber \\
&+\frac{K}{8}\left( |\eta_{ij}|^2 + |D_{ij}^a|^2 - |D_{ij}^b|^2 - |D_{ij}^{c} |^2 \right),
\label{eq:EKf}
\end{align}
where only the spin liquid channels are considered (possible magnetic channels will be considered later).
The variational wavefunction minimizing Eq.~(\ref{eq:EKf}) can be found from the self-consistent solutions of the following mean-field Hamiltonian $\widetilde{H}_K^{MF}$ such that $\langle \widetilde{H}_K^{MF} \rangle = \langle H_K \rangle$:
\begin{align}
&\widetilde{H}_K^{MF}= \sum_{a\text{-link}} \frac{K}{8} \left( \chi_{ij}^* \hat{\chi}_{ij} + E_{ij}^{a*}\hat{E}_{ij}^a -E_{ij}^{b*} \hat{E}_{ij}^b - E_{ij}^{c*} \hat{E}_{ij}^c \right) \nonumber \\
&+ \frac{K}{8}  \left( \eta_{ij}^* \hat{\eta}_{ij} + D_{ij}^{a*}\hat{D}_{ij}^a -D_{ij}^{b*} \hat{D}_{ij}^b - D_{ij}^{c*} \hat{D}_{ij}^c \right) +\mathrm{h.c.} \nonumber \\
&-\frac{K}{8}\left( |\chi_{ij}|^2 + |E_{ij}^a|^2 - |E_{ij}^b|^2 - |E_{ij}^{c} |^2 \right) \nonumber \\
&-\frac{K}{8} \left( |\eta_{ij}|^2 + |D_{ij}^a|^2 - |D_{ij}^b|^2 - |D_{ij}^{c} |^2 \right).
\label{eq:HKmf2}
\end{align}

When we impose the self-consistency on Eq.~(\ref{eq:HKmf2}), we again find the same non-vanishing mean-field parameters in Eqs.~(\ref{eq:chiz}) and (\ref{eq:Dxz}).
This means that the ground state of $H_K^{MF}$, derived from Majorana fermion mean-field theory, is also the ground state of the $\widetilde{H}_K^{MF}$ minimizing Eq.~(\ref{eq:EKf}).
However, $\langle H_K \rangle $ turns out to be four times smaller than the exact ground state energy.
The bandwidth of the gapless Majorana fermions with Eq.~(\ref{eq:HKmf2}) is also four times smaller. 
This factor of four difference is an artifact of the naive complex fermion mean-field theory in Eq.~(\ref{eq:HKmf2}); because we are imposing the one-spinon-per-site constraint only \emph{on average}, the mean-field ground state of $\widetilde{H}_K^{MF}$ does not reproduce the exact ground state energy.
On the other hand, notice that the complex fermion mean-field Hamiltonian Eq.~(\ref{eq:HKzmf}) that is obtained by performing the basis change on the Majorana mean-field theory gives the exact ground state energy, namely $E_0 = \langle H^{MF}_K\rangle = 4 \langle \widetilde{H}_K^{MF} \rangle = 4\langle H_K \rangle$. 

To elaborate on this point, let us recall that the Abrikosov fermion representation of the spin operator can be rewritten in terms of the Majorana fermions as
\begin{equation}
S_i^a = \frac{1}{2}f_{i\alpha}^\dagger \tau^a_{\alpha\beta} f_{i\beta} = \frac{1}{4} \left( i \gamma_i^a \gamma^0_i - i\gamma^b_i \gamma^c_i \right).
\label{eq:AbrikosovSpin}
\end{equation}
Because of the constraint $\gamma_i^x\gamma_i^y\gamma_i^z \gamma_i^0 = 1$, $i\gamma_i^a\gamma_i^0 = - i\gamma_i^b\gamma_i^c$ follows, and subsequently Eq.~(\ref{eq:AbrikosovSpin}) is physically equivalent to $S_i^a = \frac{i}{2}\gamma_i^a \gamma_i^0$ after projection. However, the ground state of $H_K^{MF}$ has finite correlations only for $\langle i \gamma_i^0 \gamma_j^0 \rangle$ and $\langle i \gamma_i^a \gamma_j^a \rangle$ at the $a\text{-link}$, and the other correlations $\langle i \gamma_i^\mu \gamma_j^\nu \rangle$ vanish. Therefore, only the first term of
\begin{multline}
S_i^a S_j^a = \frac{1}{4} \left[ \frac{1}{4}(i\gamma_i^a \gamma_i^0)(i\gamma_j^a \gamma_j^0) +\frac{1}{4} (i\gamma_i^b \gamma_i^c)(i\gamma_i^b \gamma_i^c) \right. \\
\left. -\frac{1}{4}(i\gamma_i^a \gamma_i^0)(i\gamma_i^b \gamma_i^c)-\frac{1}{4}(i\gamma_i^b \gamma_i^c)(i\gamma_i^a \gamma_i^0)\right]
\end{multline}
would contribute when $\langle H_K \rangle$ is evaluated, and the remaining three terms give zero energy.
However, the energy $\langle H_K^{MF} \rangle$ deduced from the Majorana fermion representation can reproduce the exact ground state energy because we fully utilized the equivalence $i\gamma_i^a \gamma_i^0 = -i\gamma_i^b \gamma_i^c$ before the mean-field decoupling is performed.
Hence, the coefficient of $H_K^{MF}$ in Eq.~(\ref{eq:HKzmf}) is four times larger than the overall coefficient of $\widetilde{H}_K^{MF}$ in Eq.~(\ref{eq:HKmf2}).

In short, if we use the mean-field Hamiltonian $\widetilde{H}_K^{MF}$ derived directly from the Abrikosov fermion representation and evaluate $ \langle H_K \rangle = \langle \widetilde{H}_K^{MF} \rangle$, the energy from the spin liquid channel would be four times smaller than the exact ground state energy.
On the other hand, the exact ground state energy can be obtained from the Majorana fermion mean-field theory in Eq.~(\ref{eq:HKmf}) or the basis-transformed Abrikosov fermion representation in Eq.~(\ref{eq:HKzmf}).

The difference in ground state energy between two representations is a result of the redundancy in parton (or slave-particle) representations of the same physical Hamiltonian.
In general, two-spin interactions are represented as four-parton interactions and the constraint allows one to rewrite the quartic interactions in various different forms.
Mean-field decoupling of such interactions leads to different representations of the mean-field Hamiltonian. 
When the constraint is exactly imposed, all of these representations give the same energy for the same ground state. 
At the mean-field level, however, they may give different energies for the same ground state.
Here, in order to reproduce the correct ground state energy in the Kitaev limit, we choose the unitary-rotated form of the Majorana fermion mean-field theory  $H_K^{MF}$ instead of the naive Abrikosov mean-field theory $\widetilde{H}_K^{MF}$.
We also keep the maximum possible number of mean-field channels for subsequent analyses. Taking into account the overall factor four difference mentioned earlier, this is equivalent to using $ \widetilde{H}_{4K}^{MF} $ with the renormalized Kitaev coupling, $K\rightarrow 4K$.

\subsubsection{Magnetic channel}

Although the pure Kitaev model has the non-magnetic, spin liquid ground state, sufficiently large coupling between the itinerant electrons and local moments may allow possible magnetic order in the Kondo-Kitaev model. 
In the Majorana representation of the spin, $S_i^a = \frac{i}{2} \gamma_i^a \gamma_i^0$, the magnetic channel can be written as
\begin{multline}
H_K^{\rm mag} = -\frac{K}{4} \sum_{a\text{-link}}
\langle i \gamma_i^a \gamma_i^0 \rangle  i \gamma_j^a \gamma_j^0 + \langle i \gamma_j^a \gamma_j^0 \rangle  i \gamma_i^a \gamma_i^0 \\
-\langle i \gamma_i^a \gamma_i^0 \rangle \langle  i \gamma_j^a \gamma_j^0 \rangle.
\label{eq:majmag}
\end{multline}
With $f_\uparrow = \frac{1}{2} \left(\gamma^0 - i \gamma^z \right)$ and $f_\downarrow = \frac{1}{2} \left( \gamma^y - i \gamma^x \right)$, we can rewrite Eq.~(\ref{eq:majmag}) in terms of the complex fermions.
Together with the one-spinon-per-site constraint, the magnetic channel in terms of $f$ fermions is
\begin{align}
H_K^{\rm mag} &=- \frac{K}{2} \sum_{a\text{-link}} \left[ s_i^a ( f_{j\alpha}^\dagger \tau^a_{\alpha\beta} f_{j\beta} ) + s_j^a ( f_{i\alpha}^\dagger \tau^a_{\alpha\beta} f_{i\beta} ) \right] \nonumber \\
&+K s_i^a s_j^a,
\end{align}
where $s_i^a = \frac{1}{2} \langle f_{i\alpha}^\dagger \tau^a_{\alpha\beta} f_{i\beta} \rangle $.

In the case of magnetic channels, the one-spinon-per-site constraint on average is sufficient to introduce the magnetic order parameters consistently.
Because $\langle f_{i\alpha}^\dagger f_{i\alpha} \rangle = 1$ and $\langle f_{i\uparrow} f_{i\downarrow} \rangle = 0$ already imply $\langle i \gamma_i^a \gamma_i^0 \rangle = - \langle i \gamma_i^b \gamma_i^c \rangle$, every physically equivalent magnetic order parameter contributes equally to the energy, e.g., $\langle S_i^a \rangle = \langle \frac{1}{4} \left( i \gamma_i^a \gamma^0_i - i\gamma^b_i \gamma^c_i \right) \rangle = \langle \frac{i}{2}\gamma_i^a \gamma_i^0 \rangle$.
Therefore we can safely introduce the magnetic channels with any representation of the spin; no enhancement or suppression of the magnetic channel is necessary in order to match the Majorana fermion mean-field theory and the Abrikosov fermion mean-field theory.

To sum up, the full mean-field Hamiltonian for the Kitaev interaction is given by
\begin{align}
&H_f= \sum_{a\text{-link}} \frac{K}{2} \left( \chi_{ij}^* \hat{\chi}_{ij} + E_{ij}^{a*}\hat{E}_{ij}^a -E_{ij}^{b*} \hat{E}_{ij}^b - E_{ij}^{c*} \hat{E}_{ij}^c \right) \nonumber \\
&+ \frac{K}{2}  \left( \eta_{ij}^* \hat{\eta}_{ij} + D_{ij}^{a*}\hat{D}_{ij}^a -D_{ij}^{b*} \hat{D}_{ij}^b - D_{ij}^{c*} \hat{D}_{ij}^c \right) +\mathrm{h.c.} \nonumber \\
&- \frac{K}{2} \left[ s_i^a ( f_{j\alpha}^\dagger \tau^a_{\alpha\beta} f_{j\beta} ) + s_j^a ( f_{i\alpha}^\dagger \tau^a_{\alpha\beta} f_{i\beta} ) \right] \nonumber \\
&-\frac{K}{2}\left( |\chi_{ij}|^2 + |E_{ij}^a|^2 - |E_{ij}^b|^2 - |E_{ij}^{c} |^2 \right) \nonumber \\
&-\frac{K}{2} \left( |\eta_{ij}|^2 + |D_{ij}^a|^2 - |D_{ij}^b|^2 - |D_{ij}^{c} |^2 \right) + K s_i^a s_j^a \nonumber \\
&+\sum_i \left[(a^x_i- i a^y_i)f_{i\downarrow} f_{i\uparrow} + \mathrm{h.c.}\right] + a^z_i (f_{i\sigma}^\dagger f_{i\sigma} -1)
\label{eq:Hf}
\end{align}
with the Lagrange multipliers $a^{x,y,z}_i$ imposing the constraints on average.
We reiterate that this Abrikosov fermion mean-field Hamiltonian is obtained by performing the unitary rotation on the Majorana fermion mean-field Hamiltonian with both spin liquid and magnetic channels.
In comparison to the naive mean-field decoupling scheme of the Abrikosov fermion representation, this Hamiltonian has four times larger weight\cite{Ogata_GutzwillerCounting, Edegger_GutzwillerRenormalization} in the spin liquid channels.
This mean-field Hamiltonian gives the spin liquid ground state and the correct ground state energy for the pure Kitaev model. 
On the other hand, if the naive decoupling scheme in the Abrikosov fermion representation were used, the spin liquid channels would have been under-estimated and the self-consistent mean-field theory would conclude that a magnetically ordered state is energetically more favourable than the spin liquid state for the pure Kitaev model.
Thus we will use the mean-field theory in Eq.(\ref{eq:Hf}), which is consistent with the exact solution, for the subsequent analysis.

\subsection{The Kondo coupling}

For the Kondo coupling term, we consider the most general mean-field Hamiltonian based on the Abrikosov fermion representation of the spin, $S_i^a =\frac{1}{2} f_{i\alpha}^\dagger \tau^a_{\alpha\beta} f_{i\beta}$.
The mean-field theory of the Kondo effect based on Abrikosov fermions reproduces exactly the exponential dependence of the Kondo temperature as function of the Kondo coupling $J_K$ \cite{Hewson}.
After we represent the spin operator in terms of $f$ fermions, we can compute the expectation value of $H_{\rm Kondo}$ using Wick's theorem.
Our mean-field Hamiltonian $H_{cf}$ is chosen such that $\langle H_{cf} \rangle = \langle H_{\rm Kondo} \rangle$.
Then, we have
\begin{align}
&H_{cf} = \sum_{i} \sum_{a=x,y,z} \frac{J_K}{8} \left ( A_{i}^{a*} \hat{A}_{i}^a + B_{i}^{a*}\hat{B}^a_{i} \right) + \mathrm{h.c.} \nonumber \\
&- \sum_i \frac{3J_K}{8} \left (A_{i}^{0*} \hat{A}^0_{i} + B_{i}^{0*} \hat{B}^{0}_{i} \right) + \mathrm{h.c.} \nonumber \\
&+\sum_i  \frac{J_K}{2}\left[ \vec{m}_i \cdot ( f_{i\alpha}^\dagger \vec{\tau}_{\alpha\beta} f_{i\beta} ) + \vec{s}_i \cdot ( c_{i\alpha}^\dagger \vec{\tau}_{\alpha\beta} c_{i\beta} )\right ] \nonumber \\
&-\sum_i \sum_{a=x,y,z} \frac{J_K}{8} \left( |A_i^a|^2 + |B_i^a|^2\right) \nonumber \\
&+\sum_i \frac{3J_K}{8} \left( |A_i^0 |^2 +|B_i^0|^2 \right) -\sum_i J_K \, \vec{m}_i \cdot \vec{s}_i,
\label{eq:Hcf}
\end{align}
where
\begin{align}
&A^0_{i} = \langle \hat{A}_{i}^0 \rangle = \langle c_{i\sigma}^\dagger f_{i\sigma} \rangle, \\
&A_{i}^a = \langle \hat{A}^a_{i} \rangle = \langle c_{i\alpha}^\dagger \tau^a_{\alpha\beta} f_{i\beta} \rangle, \\
&B^0_{i} = \langle \hat{B}^0_{i} \rangle = \langle c_{i\alpha} (-i\tau^y)_{\alpha\beta} f_{i\beta} \rangle, \\
&B_{i}^a = \langle \hat{B}^a_{i} \rangle = \langle c_{i\alpha} (i\tau^y \tau^a)_{\alpha\beta}f_{i\beta} \rangle,  \\
&m_i^a = \frac{1}{2} \langle c_{i\alpha}^\dagger \tau^a_{\alpha\beta} c_{i\beta} \rangle, \quad s_i^a = \frac{1}{2} \langle f_{i\alpha}^\dagger \tau^a_{\alpha\beta} f_{i\beta} \rangle.
\end{align}

\subsection{Projective symmetry group}

When we represent the quantum spin $\vec{S}_i$ with the fractionalized degrees of freedom $f_{i\sigma}$, we enlarge the dimension of the Hilbert space from $2^{N_\mathrm{site}}$ to $4^{N_\mathrm{site}}$.
So the spin operator can be faithfully represented with partons only if we carefully take account the constraint, $f_{i\alpha}^\dagger f_{i\alpha}=1$.
Such constrained dynamics of the strongly correlated system naturally introduces the gauge redundancy to the interacting parton Hamiltonian\cite{Nagaosa_strong, Wen_book}.

With a matrix of fermion operators\cite{Affleck_infiniteU, You_dopedJK, Huang_hyperkagome},
\begin{equation}
F_i =
\begin{pmatrix}
f_{i\uparrow} & f_{i\downarrow}^\dagger \\[3pt]
f_{i\downarrow} & -f_{i\uparrow}^\dagger,
\end{pmatrix}, \label{eq:Fmatrix}
\end{equation}
we can write the spin operator as
\begin{equation}
S^a_i = \frac{1}{4} \mathrm{Tr}(F_i^\dagger \tau^a F_i)
\end{equation}
and the one-spinon-per-site constraint as
\begin{equation}
K_i^a = \frac{1}{2} \mathrm{Tr}(F_i \tau^a F_i^\dagger) = 0.
\end{equation}
While the left $SU(2)$ rotation $F_i \rightarrow R^\dagger F_i$ leads to usual $SO(3)$ spin rotation $S_i^a \rightarrow O^{ab}S_i^b$ generated by the spin itself, the right $SU(2)$ rotation $F_i \rightarrow F_i W_i$ does not change the spin operator due to cyclic property of trace.
Therefore the parton Hamiltonian has the $SU(2)$ gauge redundancy $W_i$ generated by $K_i^a$.

Because of this $SU(2)$ gauge redundancy, there are many seemingly distinct but gauge-equivalent mean-field Hamiltonians.
To be specific, consider $H_f$ and $H_{cf}$ in terms of matrices of fermion operators:
\begin{align}
H_f &= \sum_{ij} \mathrm{Tr} \left( \tau^\alpha F_i U_{ij}^\alpha F_j^\dagger \right), \label{eq:Hfmatrix}\\
H_{cf} &= \sum_i \mathrm{Tr} \left(\tau^\alpha C_i V_i^\alpha F_i^\dagger \right), \label{eq:Hcfmatrix}
\end{align}
where $C_i$ is a matrix of $c_{i\alpha}$ fermions analogous to Eq.~(\ref{eq:Fmatrix}), and $U_{ij}^\alpha$ and $V_i^\alpha$ for $\alpha = 0,x,y,z$ are matrices of the mean-field parameters.
(Precise expressions for $U_{ij}^\alpha$ and $V_i^\alpha$ are in Appendix \ref{app:MFmatrix}.)
Under the gauge transformations $F_i \rightarrow F_i W_i$,
\begin{align}
U_{ij}^\alpha &\rightarrow W_i^\dagger U_{ij}^\alpha W_j, \\
V_i^\alpha &\rightarrow V_i^\alpha W_i
\end{align}
give the gauge-equivalent mean-field Hamiltonian.

Since one physical Hamiltonian corresponds to many gauge-equivalent mean-field Hamiltonians, the symmetry transformations do not need to preserve the structure of individual mean-field Hamiltonians.
As long as a transformation maps one mean-field Hamiltonian to the other gauge-equivalent one, the transformation is a symmetry. 
So the mean-field Hamiltonian is invariant under the symmetry transformations followed by the associated gauge transformations.
Therefore the symmetry transformation $g$ acting on the parton has the form
\begin{equation}
F_i \xrightarrow{G_g g} R_g^\dagger F_{g(i)} G_g (g(i)),
\end{equation}
where $U_g$ and $G_g(i)$ are the $SU(2)$ spin rotation and the gauge transformation associated with the symmetry $g$, respectively.
This group of symmetry transformations augmented with the gauge transformations is called the \emph{projective symmetry group} (PSG) \cite{Wen_PSG}.

An important subset of the PSG is the set of identity transformations called the invariant gauge group (IGG).
This set of pure gauge transformations characterize the emergent gauge symmetry of the low-energy physics.
For example, the $\mathbb{Z}_2$ spin liquid such as the Kitaev spin liquid has the invariant gauge group $\mathbb{Z}_2 = \{\tau^0,-\tau^0 \}$ while the trivial phase has the $\mathrm{IGG} = \{ \tau^0 \}$.

Because of this low-energy gauge redundancy, any non-trivial sequence of symmetry operations equivalent to the identity, e.g., $g=g_1 \cdot g_2 \cdot ... \cdot g_n = e$, can leave the parton invariant up to some element of the invariant gauge group: $F_i \xrightarrow{G_g g} F_i \eta_g$ where $\eta_g \in \mathrm{IGG}$.
If the sequence $g$ leaves the non-trivial pure gauge transformation $\eta_g \neq \tau^0$, those symmetries are said to be non-trivial projective symmetries.
These non-trivial projective symmetries characterize the quantum order of the ground state.

\section{Fractionalized Fermi liquid} \label{sec:FLstar}
\begin{figure}[t]
\centering
\includegraphics[width=0.49\textwidth]{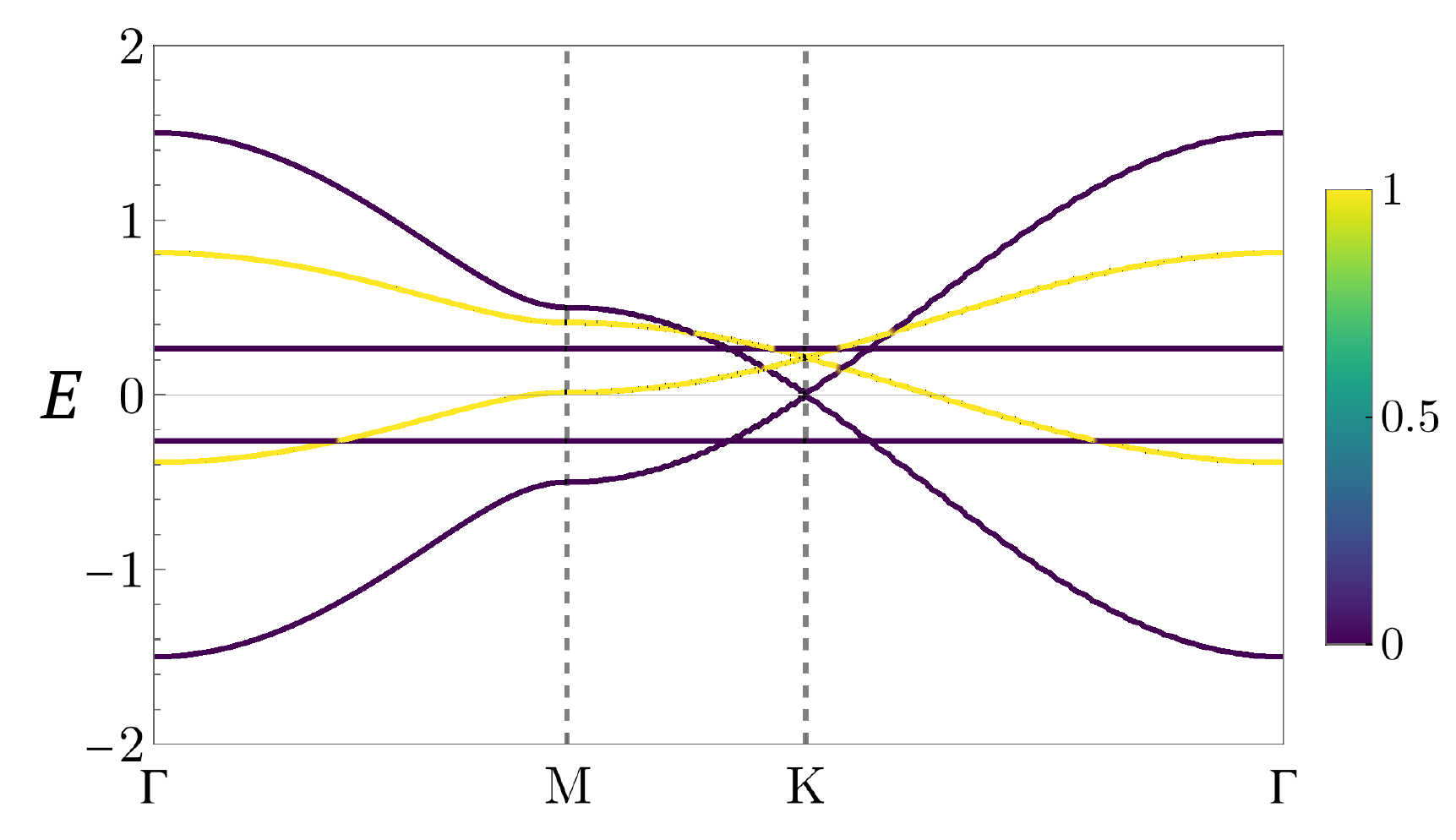}
\caption{Spectrum of the fractionalized Fermi liquid (FL*) with $t/K = 0.2$, $n_c = 0.7$.
The Majorana fermions (purple) fractionalized from the interacting spins coexist with the itinerant electrons (yellow) having ``small" Fermi surface.}
\label{fig:FLstar}
\end{figure}

In the absence of the Kondo coupling, the system is the exactly solvable Kitaev model with decoupled free electrons.
The spins are fractionalized into the Majorana fermions \cite{Kitaev_Honeycomb}, and only itinerant electrons participate in the formation of the Fermi surface.

Because the Kitaev spin liquid is stable under weak local perturbations, the spins remain fractionalized even if the itinerant electrons and the spins are weakly coupled.
With small enough Kondo coupling $J_K$, the self-consistent mean-field theory gives vanishing hybridization amplitudes and finite $f$ fermion hopping and pairing amplitudes consistent with the exact solution of the Kitaev model\cite{Schaffer_JKmodel},
\begin{align*}
&\chi_x=\chi_y=\chi_z=D_x^x = -iD_y^y=E_z^z = -0.118784i, \\
&E_x^z=-iD_x^y = E_y^z = D_y^x=D_z^x=-iD_z^y = 0.381216i,
\end{align*}
where the subscript denotes the type of the link, e.g., $E_z^x = E_{ij}^x$ for $\langle ij \rangle \in z\text{-link}$.
The Fermi surface remains ``small" in the sense that its size is determined solely from the electron filling fraction, $n_c$ (FIG. \ref{fig:FLstar}).

This \emph{fractionalized} Fermi liquid (FL*) phase does not break any symmetry, and also the projective symmetry properties of the interacting spins are preserved\cite{Senthil_FL*}.
The quantum order of the Kitaev spin liquid persists in spite of the weak Kondo coupling, and the Majorana fermions respect the symmetries up to some gauge transformations.

The projective symmetry group of the Kitaev spin liquid can be deduced from the distinct dynamic properties of the Majorana fermions $\gamma^0$ and $\gamma^{x,y,z}$.
Because the Kitaev spin liquid has well-separated static gauge degrees of freedom $u_{ij} = i\gamma^a_i \gamma^a_j $ and the gapless dynamic matter field $\gamma_i^0$, the symmetry transformations should not mix $\gamma^0$ with the other $\gamma^{x,y,z}$ fermions.
The projective extension of the symmetry group $\{ \mathcal{T}, T_{1,2}, C_6, m \}$ with the following gauge transformations is consistent with this unique property of the Kitaev spin liquid\cite{You_dopedJK}:
\begin{align}
&G_\mathcal{T}(s=0) = -G_\mathcal{T}(s=1) = i\tau^y, \label{eq:GT} \\
&G_{T_1}(i) = G_{T_2}(i) = \tau^0, \\
&G_m(s=0)  = -G_m(s=1) = R_m, \\
&G_{C_6}(s=0) = -G_{C_6}(s=1) = R_{C_6}.
\end{align}
where $s=0, 1$ are two different sublattice sites of the honeycomb lattice, and $R_m$ and $R_{C_6}$ are defined in Eqs.~(\ref{eq:Um}) and (\ref{eq:UC6}).

While we can gauge away the $G_m$ of the mirror reflection, no $SU(2)$ gauge transformation can make $G_\mathcal{T}$ and $G_{C_6}$ to be the identity matrix.
Because both $\mathcal{T}^2=e$ and $(C_6)^6=e$ leave the non-trivial $\mathbb{Z}_2$ IGG element after the projective symmetry transformations, $F_i \xrightarrow{(G_\mathcal{T} \mathcal{T})^2} -F_i$ and $F_i \xrightarrow{(G_{C_6} C_6)^6} -F_i$, the Kitaev spin liquid has the non-trivial projective time-reversal $\mathcal{T}$ and $C_6$ rotation symmetries.

An important consequence of the non-trivial projective symmetry is the robust gapless Dirac cone.
The projective time-reversal symmetry $F_i \xrightarrow{G_\mathcal{T} \mathcal{T}} \mathcal{K} i\tau^y F_i G_\mathcal{T}(i) \mathcal{K}$ requires the mean-field parameter matrix $U_{ij}^\alpha$ in Eq.~(\ref{eq:Hfmatrix}) to satisfy\cite{Huang_hyperkagome}
\begin{multline}
\sum_{ij} \mathrm{Tr}\left[ \left(\tau^y \tau^{\alpha*} \tau^y\right) F_i  G_\mathcal{T}(i) \tau^y
\left( \tau^y U_{ij}^{\alpha*}  \tau^y \right) \tau^y G_\mathcal{T}(j)^\dagger F_j^\dagger \right] \\
= - \sum_{ij} \mathrm{Tr} \left[ \tau^\alpha F_i \left( G_\mathcal{T}(i) \tau^y U_{ij}^\alpha \tau^y G_\mathcal{T}(j)^\dagger \right) F_j^\dagger \right] = H_f.
\end{multline}
Therefore
\begin{align}
G_\mathcal{T}(i)(-i\tau^y) U_{ij}^\alpha (i\tau^y)G_\mathcal{T}(j)^\dagger = -U_{ij}^\alpha.
\label{eq:Tconstraint}
\end{align}
With the gauge transformations $G_\mathcal{T} (i)$ in Eq.~(\ref{eq:GT}), the projective time-reversal symmetry does not allow any hopping or pairing of $f$ fermions between the same sublattice sites.
Hence, the mean-field Hamiltonian gains the sublattice site symmetry (or chiral symmetry) and belongs to the BDI class of the ten-fold way, where the codimension $p=d-d_{FS}=2$ Fermi surface is topologically protected\cite{Matsuura_gaplessTopological, Chiu_tenfoldway}.
Therefore small symmetry-preserving perturbations cannot gap out the Dirac cones in two spatial dimensions.

\section{Topological superconductors with the Kondo hybridization} \label{sec:SC}

When the electron-spin interaction $J_K$ is strong enough, the electrons and the spins begin to hybridize.
Because of the finite pairing amplitude of the $f$ fermions, the Kondo hybridization naturally induces the pairing of the $c$ fermions which results in the superconducting phase\cite{Senthil_FL*,SCfromFL*}.
The self-consistent mean-field theory shows that there are two phase transitions to the superconductors as we increase the Kondo coupling $J_K$: discontinuous phase transition from the $\mathbb{Z}_2$ FL* to a ferromagnetic chiral topological superconductor (FM-TSC), and continuous topological phase transition from the FM-TSC to a paramagnetic $\mathbb{Z}_2$ topological superconductor (PM-TSC).
In this section, we discuss the nature of these hybridized superconducting phases.

\subsection{Ferromagnetic topological superconductor}

\begin{figure}[t]
\centering
\includegraphics[width=0.49\textwidth]{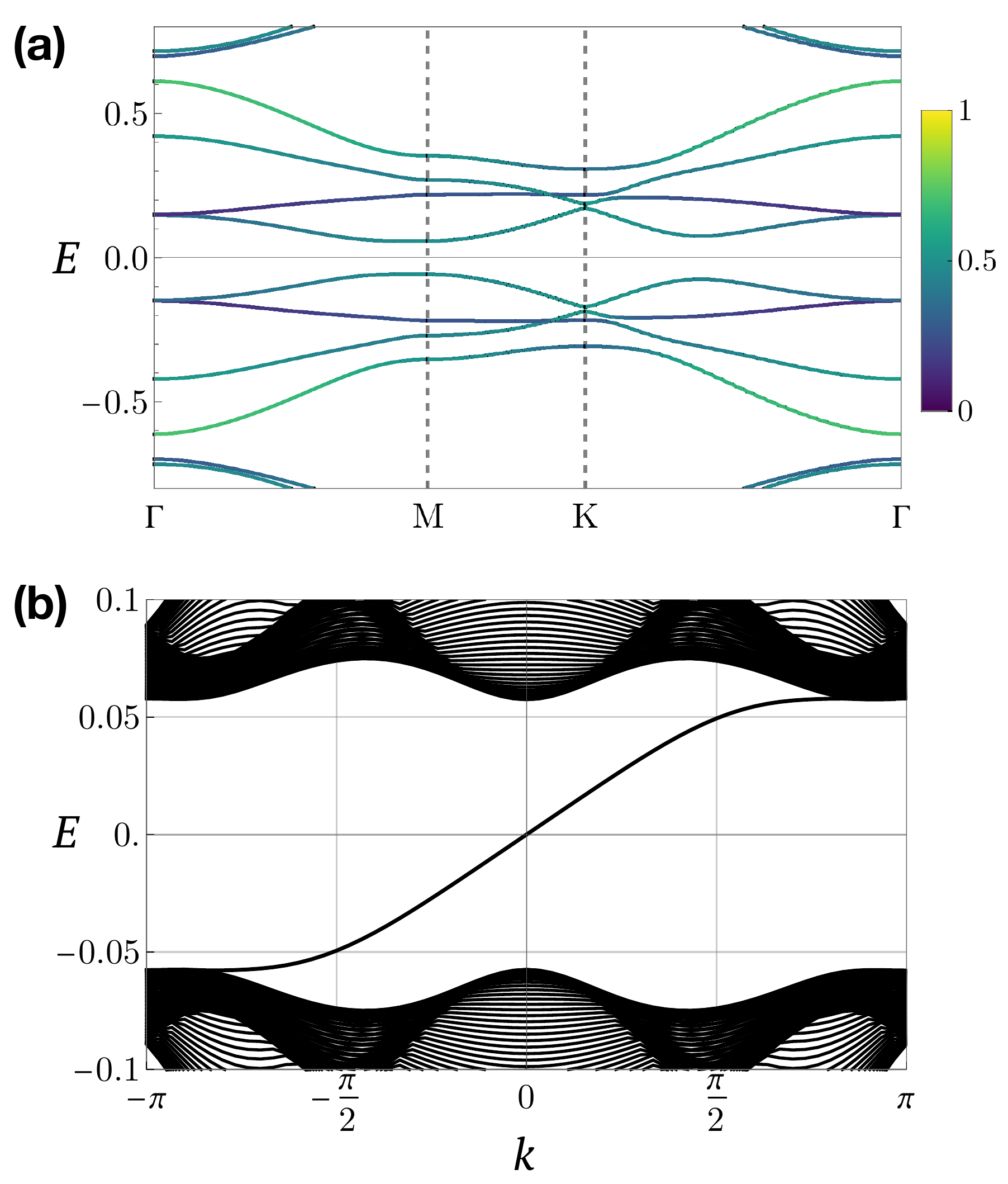}
\caption{Ferromagnetic chiral topological superconductor with $t/K=0.2$, $J_K/K=2$, and $n_c=0.7$.
(a) The excitation spectrum is fully gapped and has the total spectral Chern number, ChN = 1.
(b) Because of the non-trivial bulk topology, a stable chiral Majorana mode exists at the boundary.}
\label{fig:FM_TSC}
\end{figure}
If the electron-spinon pair is condensed under strong Kondo coupling, e.g., $\langle \hat{A}_i^0 \rangle = \langle c_{i\alpha}^\dagger f_{i\alpha} \rangle \neq 0$, the fractionalized Fermi liquid phase is destabilized and becomes a ferromagnetic topological superconductor (FM-TSC).
Because $H_{cf}$ is not invariant under the global sign flip of the $f$ fermions, the invariant gauge group of the $f$ fermions is reduced from $\mathbb{Z}_2 = \{ \tau^0, -\tau^0 \}$ to the trivial group $\{\tau^0 \}$.
In other words, the superconducting phase no longer possesses the emergent $\mathbb{Z}_2$ gauge symmetry for the spins\cite{Senthil_FL*,SCfromFL*}.
Therefore the notion of the quantum order/non-trivial projective symmetry no longer exists in this hybridized superconducting phase.

The breakdown of the PSG has important consequences for the symmetry properties of the FM-TSC phase.
Since the FL* phase has the non-trivial projective time-reversal $\mathcal{T}$ and $C_6$ rotation symmetry, the lack of the gauge structure may lead to the spontaneous symmetry breaking\cite{Rau_2014} of $\mathcal{T}$ and $C_6$.
Because the mean-field parameters for hopping and pairing change discontinuously right after the phase transition, the self-consistent mean-field parameters are adjusted as
\begin{align}
&\chi_x = \chi_y = \chi_z \in \mathbb{R}, \label{eq:FMchi}\\
&\eta_x = \eta_y = \eta_z \in \mathbb{R}, \\
&E^x_x = E^y_y = E^z_z \in \mathbb{R}, \\
&E^y_x = E^{z*}_x = E^z_y = E^{x*}_y = E^x_z = E^{y*}_z \in \mathbb{C}, \\
&D^x_x = D^y_y = D^z_z \in \mathbb{R}, \\
&D^y_x = D^{z*}_x = D^z_y = D^{x*}_y = D^x_z = D^{y*}_z\in \mathbb{C},  \\
&A^0_{s=0,1} \in \mathbb{R}, \quad B^0_{s=0,1} \in \mathbb{R}, \\
&A^x_{s=0,1} = A^y_{s=0,1} = A^z_{s=0,1} \in \mathbb{R}, \\
&B^x_{s=0,1} = B^y_{s=0,1} = B^z_{s=0,1} \in \mathbb{R}, \label{eq:FMB}
\end{align}
such that the hybridized superconducting phase could preserve the $C_3$ rotation symmetry; $H_{MF} = H_c + H_f + H_{cf}$ is invariant under $F_i \rightarrow (R_{C_6}^\dagger)^2 F_{C_3(i)}$ and $C_i \rightarrow C_{C_3(i)}$.
In general, the self-consistent mean-field parameters do not have the structure of Eqs.~(\ref{eq:FMchi})--(\ref{eq:FMB}) because of the $SU(2)$ gauge redundancy $F_i \xrightarrow{g} R_g^\dagger F_{g(i)} \xrightarrow{G_g} R_g^\dagger F_{g(i)} G_g(g(i))$.
However, we can always choose the gauge such that $G_g(i) = \tau^0$ for every symmetry $g$ because the finite Kondo hybridization amplitudes do not allow any non-trivial projective symmetry \footnote{The $SU(2)$ gauge redundancy is intrinsic nature of the Abrikosov fermion representation.
The breakdown of the PSG means the trivial invariant gauge group and does not imply lack of the $SU(2)$ gauge redundancy.}. Note that non-trivial projective symmetries $\mathcal{T}$ and $C_6$ prevent such gauge choice for the FL* phase. All discussion from now on assumes the gauge such that $G_g(i) = \tau^0$, which gives Eqs.~(\ref{eq:FMchi})--(\ref{eq:FMB}).

As we advertised earlier, the pairing correlation of $f$ fermions induces the pairing of the itinerant electrons.
The electron pairing is purely triplet, and its amplitudes $\Delta^\mu_{\nu} = \langle c_{i\alpha} (i\tau^y \tau^\mu)_{\alpha\beta} c_{j\beta} \rangle $ at the $\langle ij \rangle \in \nu\text{-link}$ are
\begin{align}
&\Delta^a_{a} \in \mathbb{R}, \quad \Delta^b_{a} = \Delta^{c*}_{a} \in \mathbb{C},\\
&\Delta_x^{x,y,z} = \Delta_y^{y,z,x} = \Delta_z^{z,x,y},
\end{align}
where $a, b, c$ are the cyclic indices for $x, y, z$.
If the state were time-reversal symmetric, the triplet pairing amplitudes must be purely imaginary.
Therefore the real component of $\Delta_{ij}^\mu$ must be a consequence of spontaneously broken time-reversal symmetry with ferromagnetic order,
\begin{align}
&m= m_{s=0,1}^x = \pm m_{s=0,1}^y = \pm m_{s=0,1}^z, \\
&s=s_{s=0,1}^x =  \pm s_{s=0,1}^y = \pm s_{s=0,1}^z.
\end{align}
Depending on the sign of the $x, y, z$ components of the magnetic order parameters $s_i^\mu$, the PM-TSC state is $2^3 = 8$ fold degenerate.
Without loss of generality, we will focus on the state with $s_i^x = s_i^y = s_i^z=s$ and $m_i^x = m_i^y = m_i^z=m$.
[Eqs.~(\ref{eq:FMchi})--(\ref{eq:FMB}) are consistent with this direction.]

When time-reversal symmetry is broken, the same sublattice site pairing makes the Kitaev spin liquid become the non-Abelian topological phase with the spectral Chern number $\pm 1$\cite{Kitaev_Honeycomb}.
The question arises whether the hybridized superconductor also inherits this topological property of the parent spin liquid.
The self-consistent mean-field theory shows that the hybridized superconductor is indeed a chiral topological superconductor (class D) with the spectral Chern number, ChN = 1.
A chiral Majorana mode at the edge manifests the non-trivial bulk topology and spontaneously broken time-reversal symmetry [FIG. \ref{fig:FM_TSC} (b)].
Based on Kitaev's 16 fold way classification of two dimensional chiral superconductors\cite{Kitaev_Honeycomb}, the vortex excitation of this chiral superconductor with odd Chern number is the Ising anyon with non-Abelian braiding statistics because of a single unpaired Majorana mode inside the vortex core\cite{Read_p+ip, Ivanov_halfVortex}.
While the FM-TSC no longer has the Ising gauge structure of the chiral spin liquid phase of the Kitaev model, nevertheless the nature of the excitations shows a surprising similarity.

\subsection{Paramagnetic topological superconductor}

\begin{figure}[t]
\centering
\includegraphics[width=0.49\textwidth]{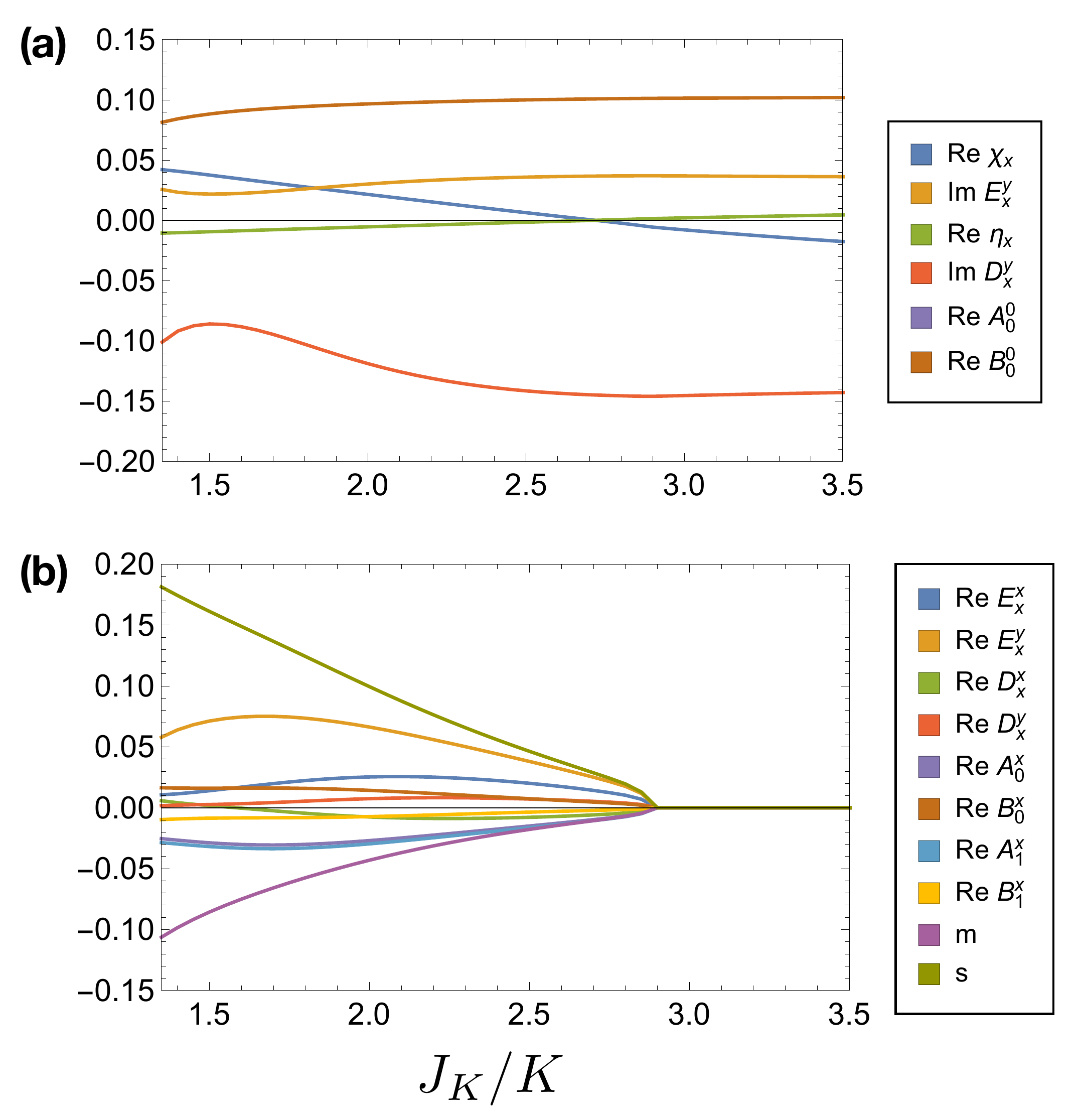}
\caption{Self-consistent mean-field parameters for the hybridized superconducting phases ($t/K=0.2$, $n_c=0.7$).
Because of the $C_3$ symmetry, only the mean-field parameters for the $x\text{-links}$ are shown.
(a) The mean-field parameters consistent with time-reversal symmetry are a continuous function of $J_K$.
(b) Time-reversal symmetry breaking mean-field parameters such as the real part of triplet hopping($E$)/pairing($D$)/hybridization($A, B$) and the magnetic order parameters ($m, s$) continuously go to zero at the critical point, $(J_K)_c/K=2.9$.}
\label{fig:MFparam}
\end{figure}

\begin{figure}[t]
\centering
\includegraphics[width=0.49\textwidth]{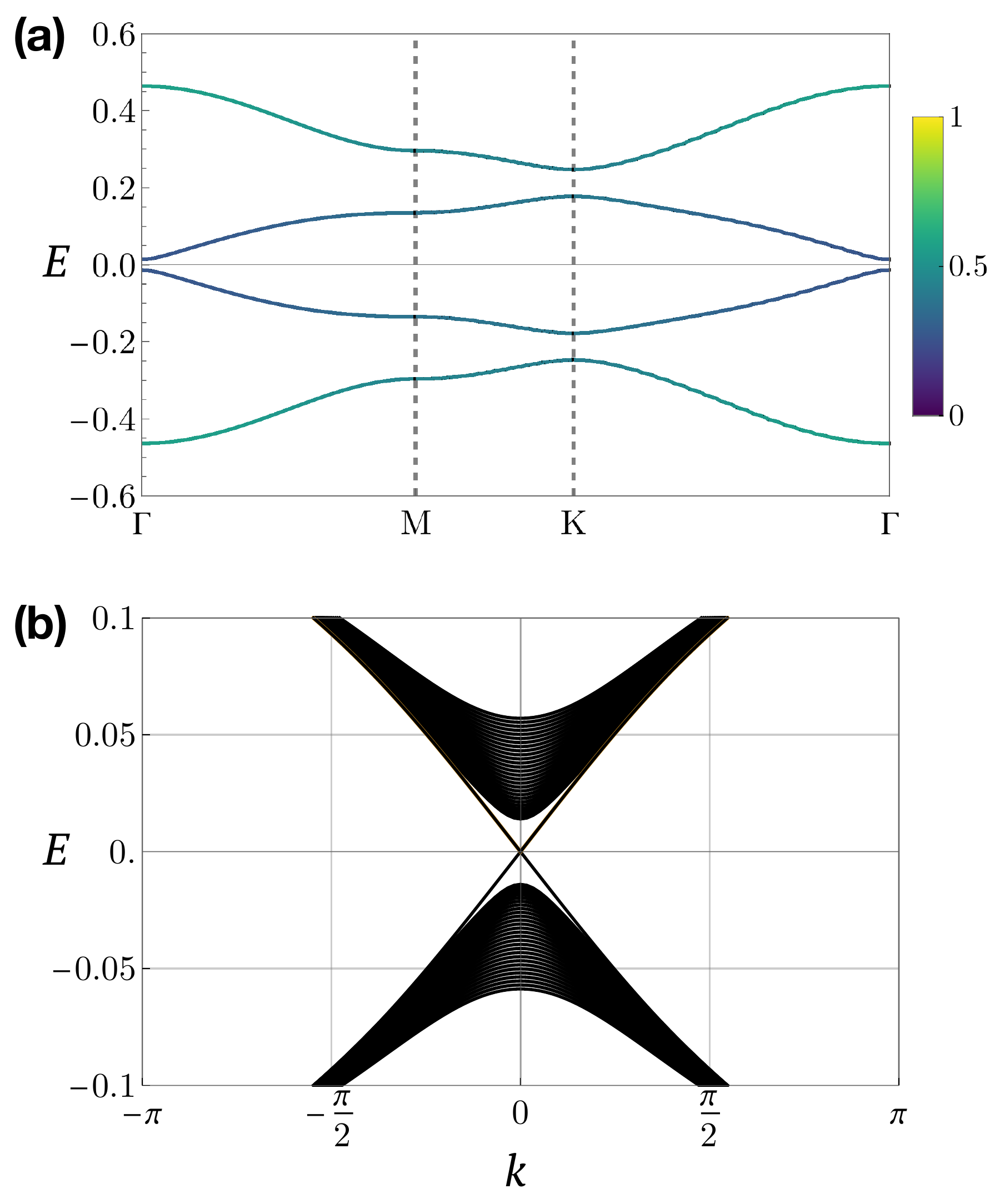}
\caption{Paramagnetic topological superconductor ($t/K=0.2$, $J_K/K=3$, $n_c=0.7$)
(a) The bulk excitation spectrum is fully gapped over the entire Brillouin zone, and all bands are doubly degenerate due to time-reversal symmetry and inversion-like discrete symmetry.
(b) A robust single helical Majorana edge mode manifests the non-trivial bulk topology and recovered time-reversal symmetry.}
\label{fig:PM_TSC}
\end{figure}

When the Kondo coupling $J_K$ becomes much stronger than the Kitaev interaction, the magnetic order parameters and all time-reversal symmetry breaking mean-field parameters continuously decrease to zero [FIG. \ref{fig:MFparam} (b)].
At the critical point, the mean-field spectrum becomes gapless at the $\Gamma$ point.
Through this continuous phase transition, our system becomes a paramagnetic topological superconductor (PM-TSC).
The ferromagnetic order is completely suppressed, and the triplet pairing amplitudes of the electrons at $a$-links are purely imaginary:
\begin{align}
&\Delta^a_{a} =0, \quad i\Delta^b_{a} =- i\Delta^{c}_{a} \in \mathbb{R},\\
&\Delta_x^{x,y,z} = \Delta_y^{y,z,x} = \Delta_z^{z,x,y}.
\end{align}
Hence, the PM-TSC is a time-reversal symmetric superconductor, which belongs to the class DIII of the ten-fold way classification.

In two spatial dimensions, a class DIII superconductor can be a $\mathbb{Z}_2$ topological superconductor\cite{Schnyder_tenfoldway, Chiu_tenfoldway}.
Indeed, our paramagnetic fully-gapped superconducting phase has a non-trivial bulk topology which results in a robust single gapless helical edge mode [FIG. \ref{fig:PM_TSC} (b)].
Because the Kitaev spin liquid's stable Majorana cones are the consequence of both the emergent $\mathbb{Z}_2$ gauge symmetry and time-reversal symmetry, the system can have fully gapped spectrum just by destroying the gauge structure by the Kondo hybridization.

In addition to the topological properties, one notable feature of the PM-TSC phase is its symmetry.
Including time-reversal symmetry, the PM-TSC recovers every space group symmetry of the physical Hamiltonian with the exception of inversion symmetry.
To examine mirror reflection and inversion symmetry, we need to investigate the structure of the self-consistent mean-field parameters.
The non-vanishing mean-field parameters for the PM-TSC are
\begin{align}
&\chi_x = \chi_y = \chi_z \in \mathbb{R}, \\
&\eta_x = \eta_y = \eta_z \in \mathbb{R}, \\
&iE^y_x = - iE^z_x = iE^z_y = - iE^x_y =  iE^x_z = - iE^y_z \in \mathbb{R}, \label{eq:Es}\\
&iD^y_x = - iD^z_x = iD^z_y = - iD^x_y =  iD^x_z = - iD^y_z \in \mathbb{R}, \label{eq:Ds} \\
&A^0_{s=0} = A^0_{s=1} = A \in \mathbb{R}, \\
&B^0_{s=0} = B^0_{s=1} = B \in \mathbb{R}.
\end{align}
Because the singlet hybridization amplitudes $A$ and $B$ are independent of the sublattice sites, $H_{cf}$ is invariant under both mirror reflection and inversion.
So the only possible source of the symmetry breaking is $H_f$.
If $H_f$ respects mirror reflection symmetry, then
\begin{align}
&H_f = \mathrm{Tr} \left( \tau^\alpha F_{i} U_{ij}^\alpha F_{j}^\dagger \right) = \mathrm{Tr} \left(R_m \tau^\alpha R_m^\dagger F_{m(i)} U_{ij}^\alpha F_{m(j)}^\dagger \right) \nonumber \\
&\Rightarrow U_z^{x}=U_z^{y\dagger},~ U_z^z = - U_z^{z\dagger},~ U_x^{x,y} = U_y^{y,x\dagger},~ U_x^z = -U_y^{z\dagger}.
\label{eq:mConstraint}
\end{align}
With the mean-field parameter matrices in the Appendix \ref{app:MFmatrix}, it is not difficult to verify that the self-consistent mean-field parameters for the PM-TSC satisfy Eq.~(\ref{eq:mConstraint}). Therefore, the paramagnetic superconductor respects mirror symmetry.

Unlike mirror reflection, physical inversion (or $C_2$ rotation around the centre of the hexagon) turns out to be not a symmetry of the PM-TSC.
Because of translation symmetry, the mean-field parameters for the nearest-neighbour link depend only on the sublattice sites $s=0,1$ and the type of the link. Hence, inversion on the lattice is equivalent to the change of the sublattice sites $U_{ij} = U_{s=0,s=1} \xrightarrow{C_2} U_{i' j'} = U_{s=1,s=0}$.
So the usual physical inversion demands $U_\mu^\nu = U_\mu^{\nu\dagger}$, which is violated with our mean-field parameters in Eqs.~(\ref{eq:Es}) and (\ref{eq:Ds}).

Even though the lattice inversion symmetry is broken, the nearest-neighbour mean-field Hamiltonian respecting translation, mirror reflection, and $C_3$ rotation symmetry has hidden inversion-like discrete symmetry, $\mathcal{P}$\footnote{Inversion-like discrete symmetry $\mathcal{P}$ is \emph{not} a composition of translation, mirror reflection, and $C_3$ rotation symmetry. Instead one can read off the hidden symmetry $\mathcal{P}$ from the analytic expression of the nearest-neighbour mean-field Hamiltonian constrained by $T_{1,2}$, $m$, and $C_3$.}.
First, let's see how $T_{1,2}$, $m$, and $C_3$ constrain the structure of the nearest-neighbour mean-field Hamiltonian. Then we will read out the hidden symmetry $\mathcal{P}$ from the mean-field ansatz constrained by those three symmetries.

For a translationally invariant mean-field ansatz $U_{ij}^\alpha$, mirror reflection constrains the $z\text{-links}$ which contain the mirror plane as $U_z^x = U_z^{y\dagger}$ and $U_z^z = -U_z^{z\dagger}$ [Eq.~(\ref{eq:mConstraint})].
Because of $C_3$ rotation symmetry, we have three mirror planes perpendicular to each bond of the honeycomb lattice.
Thus, the constraints on the mean-field parameters on the $z$-links in Eq.~(\ref{eq:mConstraint}) can be extended to all three types of links:
\begin{equation}
U_a^{b}=U_a^{c\dagger},~ U_a^a = - U_a^{a\dagger}.
\label{eq:Pconstraint}
\end{equation}
The usual physical inversion is not a symmetry because it does not flip the spins.
However, if we act the usual physical inversion ($C_2$) to the lattice \emph{and} mirror-reflect the spins, i.e.,
\begin{equation}
F_i \xrightarrow{\mathcal{P}} R_m^\dagger F_{C_2(i)},
\end{equation}
then the mean-field Hamiltonian for the PM-TSC is invariant.
In other words, when the mean-field ansatz satisfies Eq.~(\ref{eq:Pconstraint}), we get inversion-like discrete symmetry $\mathcal{P}$ for free because Eq.~(\ref{eq:Pconstraint}) is necessary and sufficient condition for
\begin{align}
H_f &= \mathrm{Tr} \left( \tau^\alpha F_{i} U_{ij}^\alpha F_{j}^\dagger \right) \nonumber \\
&\xrightarrow{\mathcal{P}} \mathrm{Tr} \left(R_m \tau^\alpha R_m^\dagger F_{C_2(i)} U_{ij}^\alpha F_{C_2(j)}^\dagger \right) = H_f.
\label{eq:Pconstraint2}
\end{align}
Since $\mathcal{P}$ transforms the lattice sites as inversion symmetry ($C_2$), $\mathcal{P}$ relates the excitation of momentum $k$ and the excitation of momentum $-k$ in Fourier space.

Because of this inversion-like discrete symmetry $\mathcal{P}$ and time-reversal symmetry $\mathcal{T}$, each band of the excitation spectrum is doubly degenerate [FIG. \ref{fig:PM_TSC} (a)].
Time-reversal symmetry implies $E_{k,s} = E_{-k,s}$ while the inversion-like discrete symmetry implies $E_{k,s} = E_{-k,1-s}$.
Therefore we have Kramer's degeneracy $E_{k,s}=E_{k,1-s}$ with antiunitary symmetry $\mathcal{PT}$ at each $k$.
Note that the two-fold degeneracy originates not from the spins, but from the two sublattice sites $s=0,1$.

\section{Conclusion} \label{sec:conc}

In this work, we identify and characterize emergent quantum phases in the Kondo-Kitaev model on the honeycomb lattice.
The self-consistent Abrikosov fermion mean-field theory gives three distinct phases as the Kondo coupling is increased: the fractionalized Fermi liquid (FL*), where the Kitaev spin liquid and conduction electrons remain decoupled, the ferromagnetic topological superconductor (FM-TSC) of Class D with broken time-reversal symmetry, and the paramagnetic topological superconductor (PM-TSC) of Class DIII with preserved time-reversal symmetry.

Because the Kitaev spin liquid respects time-reversal symmetry ($\mathcal{T}$) projectively, the discontinuous transition from the FL* to the FM-TSC accompanies spontaneously broken time-reversal symmetry.
This $\mathcal{T}$-broken hybridized superconductor has finite ferromagnetic order and fully gapped spectrum with non-trivial bulk topology.
It is reminiscent of the phase transition from the Kitaev spin liquid to a chiral spin liquid, which occurs when an external magnetic field is applied.
In fact, one can see that the FM-TSC inherits the topological properties of the chiral spin liquid, namely the unit Chern number and a single chiral Majorana edge mode, albeit the time-reversal symmetry is spontaneously broken without any explicit $\mathcal{T}$-breaking perturbation.

Further increase of the Kondo coupling allows a continuous transition to the time-reversal symmetric PM-TSC, where the ferromagnetic order and every $\mathcal{T}$-breaking mean-field channel goes to zero continuously.
This topological superconductor is characterized by $\mathbb{Z}_2$ bulk topological invariant and a single helical edge mode. 

In the current work, we focus on the possibility of topological superconductivity in the Kondo-Kitaev system. For this purpose, we presented the results of $t/K$ = 0.2 and varied strength of $J_K$ as well as the conduction electron filling $n_c = 0.7$.
While we checked that similar values of $t/K$ and $n_c$ lead to qualitatively the same phase diagram, we have not explored all the possible cases.
Obtaining the full phase diagram as a function of different conduction electron filling factors and different relative strength between $t, K, J_K$ would be an interesting, but time-consuming, exercise for future studies.

\begin{acknowledgments}
This work was supported by the NSERC of Canada and Center for Quantum Materials at the University of Toronto (WC and YBK),
and DFG project C02 of CRC1238 (PWK and AR).
PWK acknowledges support from the German Academic Scholarship Foundation and the Bonn-Cologne Graduate School of Physics and Astronomy.
\end{acknowledgments}

\appendix

\section{Mean-field parameter matrix} \label{app:MFmatrix}
With the matrix of the fermion operators,
\begin{align}
C_i &=
\begin{pmatrix}
c_{i\uparrow} & c_{i\downarrow}^\dagger \\[3pt]
c_{i\downarrow} & -c_{i\uparrow}^\dagger,
\end{pmatrix},
&
F_i &=
\begin{pmatrix}
f_{i\uparrow} & f_{i\downarrow}^\dagger \\[3pt]
f_{i\downarrow} & -f_{i\uparrow}^\dagger,
\end{pmatrix},
\end{align}
any quadratic Hamiltonian can be written in terms of $C_i$ and $F_i$. So we reexpress $H_f$ and $H_{cf}$ as
\begin{align}
H_f &= \sum_{ij} \mathrm{Tr}\left( \tau^\alpha F_i U_{ij}^\alpha F_j^\dagger \right), \\
H_{cf} &=\sum_i \mathrm{Tr}\left( \tau^\alpha C_i V_i^\alpha F_i^\dagger \right).
\end{align}

Because the Hamiltonian must be Hermitian, the matrix of the mean-field parameters $U_{ij}^\alpha$ and $V_i^\alpha$ should have the following structure:
\begin{align}
&U_{ij}^0 = i \tau^0 (u_{ij}^0)_0 + \sum_{l=x,y,z} \tau^l (u_{ij}^0)_l, \\
&U_{ij}^{x,y,z} = \tau^0 (u_{ij}^{x,y,z})_0  + \sum_{l=x,y,z} i \tau^l(u_{ij}^{x,y,z})_l , \\
&V_{i}^0 = i \tau^0  (v_{i}^0)_0+ \sum_{l=x,y,z}  \tau^l(v_{i}^0)_l, \\
&V_{i}^{x,y,z} =  \tau^0 (v_{i}^{x,y,z})_0 + \sum_{l=x,y,z} i\tau^l (v_{i}^{x,y,z})_l,
\end{align}
where $(u_{ij}^\mu)_\nu$ and $(v_{i}^\mu)_\nu$ are all real numbers.

For the $\langle ij \rangle \in a\text{-link}$, the mean-field parameter matrix $U_{ij}^\alpha \equiv U_{a}^\alpha$ are

\begin{align}
U_{a}^0 &= -\frac{K}{2}
\begin{pmatrix}
\chi_{a} & \eta_{a}^{*} \\
\eta_{a} & -\chi_{a}^{*}
\end{pmatrix}, &
U_{a}^{a} &= -\frac{K}{2}
\begin{pmatrix}
E_{a}^a & D_{a}^{a*} \\
-D_{a}^a & E_{a}^{a*}
\end{pmatrix}, \nonumber \\
U_{a}^{b} &= \frac{K}{2}
\begin{pmatrix}
E_{a}^b & D_{a}^{b*} \\
-D_{a}^b & E_{a}^{b*}
\end{pmatrix}, &
U_{a}^{c} &= \frac{K}{2}
\begin{pmatrix}
E_{a}^c & D_{a}^{c*} \\
-D_{a}^c & E_{a}^{c*}
\end{pmatrix}.
\label{eq:Umatrix}
\end{align}
The on-site terms $(i=j)$ such as the Lagrange multipliers $a_i^{x,y,z}$ and the magnetic order parameter $s_i^{x,y,z}$ are
\begin{align}
&U_i^0 =-
\begin{pmatrix}
a_i^z & a_i^x - i a_i^y \\
a_i^x + i a_i^y & -a_i^z
\end{pmatrix},
\\
&U_i^{x,y,z} = \frac{K}{2}
\begin{pmatrix}
s_j^{x,y,z} & 0 \\
0 & s_j^{x,y,z}
\end{pmatrix}
-\frac{v}{2}
\begin{pmatrix}
m_i^{x,y,z} & 0 \\
0 & m_i^{x,y,z}
\end{pmatrix}
\end{align}
Similarly, hybridization amplitudes can be organized into the matrix $V_i^\alpha$:
\begin{align}
&V_i^0 = -\frac{3J_K}{8}
\begin{pmatrix}
A_i^0 & B^{0*}_i \\
B_i^0 & -A_i^{0*}
\end{pmatrix},
\\
&V_i^{x,y,z} = \frac{J_K}{8}
\begin{pmatrix}
A_i^{x,y,z} & B^{x,y,z*}_i \\
-B_i^{x,y,z} & A_i^{x,y,z*}
\end{pmatrix}.
\end{align}

\bibliography{KondoKitaev}

\end{document}